# Phase sampling Profilometry

## Zhenzhou Wang *

*College of Electrical and Electronic Engineering,
Shandong University of Technology, Zibo, 255000, China*
*\*Corresponding author: wangzz@sdut.edu.cn*

*Abstract*—Structured light 3D surface imaging is a school of techniques in which structured light patterns are used for measuring the depth map of the object. Among all the designed structured light patterns, phase pattern has become most popular because of its high resolution and high accuracy. Accordingly, phase measuring profilometry (PMP) has become the mainstream of structured light technology. In this letter, we introduce the concept of phase sampling profilometry (PSP) that calculates the phase unambiguously in the spatial-frequency domain with only one pattern image. Therefore, PSP is capable of measuring the 3D shapes of the moving objects robustly with single-shot.

## 1 Introduction

Structured light 3D surface imaging technology has been developed to measure the 3D shapes of the objects quantitatively, which aims to provide a data basis for reverse engineering, better understanding and a variety of industrial applications[1-3]. So far, the most widely used structured light patterns are phase patterns that are generated by the sinusoidal functions. Nowadays, there are two categories of 3D surface imaging techniques that are based on the phase pattern. One is the Fourier transform profilometry[4-6] and the other is phase shifting profilometry [6-10]. The Fourier transform profilometry generates a single high-frequency phase pattern and calculates the phase based on Fourier analysis[4]. Fourier transform profilometry requires the frequency of the carrier higher frequency to be high enough to reduce the aliasing error. Because it is single-shot, Fourier transform profilometry is suitable for measuring moving objects. In contrast, the phase shifting profilometry calculates the phase in the temporal domain by projecting multiple phase patterns with deliberately designed shifts. Up to date, numerous phase shifting algorithms have been proposed for pixel-by-pixel phase calculation[6-10]. Because the measured object should be kept stationary during continuously projecting multiple phase patterns onto it, phase shifting profilometry is not suitable for measuring moving objects by nature. However, many researchers have overcome this inherent defect of phase shifting profilometry by adopting cameras and projectors with high speed frame rates to measure the dynamic scenes[7-10]. When the frame rate is much higher than the moving speed of the measured object, the error caused by the motion of the object could be neglected during calculating the phase. Both Fourier transform profilometry and phase shifting profilometry require the phase unwrapping algorithms to retrieve the natural phase, which is not a trivial job[11].

Compared to other structured light patterns[1-3], the most attractive part of phase patterns is the full-resolution, i.e. the pixel-by-pixel 3D measurement. Though many excellent single-shot structured light methods[12-16] have been proposed to measure the moving objects, phase patterns remained the most widely used patterns both in academic research and in industrial applications. One common purpose of all the single-shot structured light methods is to extract the designed structured light patterns unambiguously from only one single pattern image, which has remained as the most challenging task so far[1]. To make the pattern extractable from one single image, the resolution of the structured light pattern needs to be decreased compared to the phase pattern. After the designed pattern is extracted robustly, the 3D information could be easily measured by non-linear mapping or triangulation[1]. The reason why multiple phase patterns are required by PSP is that the phase could not be determined unambiguously with a single image. If the complete phase pattern could be extracted from a single image robustly, no phase shifting will be required any more.

In this letter, we introduce the concept of phase sampling profilometry. We utilize the Shannon sampling theory[17-18] to design a down-sampled phase pattern and recover the full-resolution phase pattern in one-dimension based on a low-pass filter in the frequency domain. After down-sampling, the phase pattern could be extracted robustly by image processing from a single image. According to Shannon sampling theorem, the full-resolution phase could be recovered completely if the sampling frequency rate is greater or equal to twice the maximum frequency of the designed phase pattern. Thus, the full-resolution phase pattern could be recovered perfectly if the pattern sampling rate is designed to meet the Shannon sampling theorem perfectly.

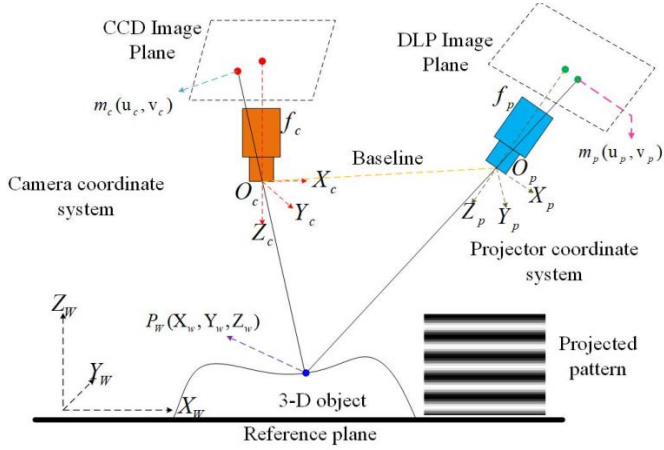

Fig. 1. The 3D imaging model of stereo vision.

## 2 THE CONCEPT OF STRUCTURED LIGHT 3D SURFACE IMAGING

Before moving to the central content of PSP, we first introduce the concept of structured light 3D imaging. As demonstrated in Fig. 1, the structured light 3D imaging system is composed of a digital light processor (DLP) projector and a charge coupled device (CCD) camera. The mathematical relationship between the camera coordinate system and the world coordinate system follows the rule of pinhole camera model that could be described by the following equation.

$$Z_c \begin{bmatrix} x_c \\ y_c \\ 1 \end{bmatrix} = M_{int} M_{ext} \begin{bmatrix} X_w \\ Y_w \\ Z_w \\ 1 \end{bmatrix} \quad (1)$$

where $(x_c, y_c)$ is the camera coordinate and $(X_w, Y_w, Z_w)$ is the world coordinate. $M_{int}$ is the intrinsic matrix and $M_{ext}$ is the extrinsic matrix that are formulated as:

$$M_{int} = \begin{bmatrix} f_x & 0 & u_c \\ 0 & f_y & v_c \\ 0 & 0 & 1 \end{bmatrix} \quad (2)$$

$$M_{ext} = \begin{bmatrix} r_{11}, r_{12}, r_{13}, T_1 \\ r_{21}, r_{22}, r_{23}, T_2 \\ r_{31}, r_{32}, r_{33}, T_3 \end{bmatrix} \quad (3)$$

where $(f_x, f_y)$ is camera's focal length and $(u_c, v_c)$ is camera's principle point. $[r_{11}, r_{12}, r_{13}; r_{21}, r_{22}, r_{23}; r_{31}, r_{32}, r_{33}]$ is the rotation matrix and $[T_1, T_2, T_3]^T$ is the translation vector. For simplicity, Eq. (1) could be rewritten as the following format.

$$Z_c \begin{bmatrix} x_c \\ y_c \\ 1 \end{bmatrix} = \begin{bmatrix} m_{11}^{wc}, m_{12}^{wc}, m_{13}^{wc}, m_{14}^{wc} \\ m_{21}^{wc}, m_{22}^{wc}, m_{23}^{wc}, m_{24}^{wc} \\ m_{31}^{wc}, m_{32}^{wc}, m_{33}^{wc}, m_{34}^{wc} \end{bmatrix} \begin{bmatrix} X_w \\ Y_w \\ Z_w \\ 1 \end{bmatrix} \quad (4)$$

To decrease the number of unknown parameters, we let $m_{34}^{wc} = 1$. Eq. (4) is then expanded into the following three equations.

$$Z_c x_c = m_{11}^{wc} X_w + m_{12}^{wc} Y_w + m_{13}^{wc} Z_w + m_{14}^{wc} \quad (5)$$

$$Z_c y_c = m_{21}^{wc} X_w + m_{22}^{wc} Y_w + m_{23}^{wc} Z_w + m_{24}^{wc} \quad (6)$$

$$Z_c = m_{31}^{wc} X_w + m_{32}^{wc} Y_w + m_{33}^{wc} Z_w + 1 \quad (7)$$

Substituting Eq. (7) into Eq. (5), we get the following equation.

$$m_{11}^{wc} X_w + m_{12}^{wc} Y_w + m_{13}^{wc} Z_w + m_{14}^{wc} = \\ x_c m_{31}^{wc} X_w + x_c m_{32}^{wc} Y_w + x_c m_{33}^{wc} Z_w + x_c \quad (8)$$

Substituting Eq. (7) into Eq. (6), we get the following equation.

$$m_{21}^{wc} X_w + m_{22}^{wc} Y_w + m_{23}^{wc} Z_w + m_{24}^{wc} = \\ y_c m_{31}^{wc} X_w + y_c m_{32}^{wc} Y_w + y_c m_{33}^{wc} Z_w + y_c \quad (9)$$

Eq.(8) could be rewritten into the following format:

$$[X_w, Y_w, Z_w, 1, 0, 0, 0, 0, -x_c X_w, -x_c Y_w, -x_c Z_w] \theta_c = x_c \quad (10)$$

where

$$\theta_c = \left[ m_{11}^{wc}, m_{12}^{wc}, m_{13}^{wc}, m_{14}^{wc}, m_{21}^{wc}, m_{22}^{wc}, m_{23}^{wc}, m_{24}^{wc}, m_{31}^{wc}, m_{32}^{wc}, m_{33}^{wc} \right]^T$$

In the same way, Eq.(9) could be rewritten into the following format:

$$[0, 0, 0, 0, X_w, Y_w, Z_w, 1, -y_c X_w, -y_c Y_w, -y_c Z_w] \theta_c = y_c \quad (11)$$

Eq.(10) and Eq. (11) are combined into the following equation.

$$X_c \theta_c = Y_c \quad (12)$$

where

$$X_c = \begin{bmatrix} X_w, Y_w, Z_w, 1, 0, 0, 0, 0, -x_c X_w, -x_c Y_w, -x_c Z_w \\ 0, 0, 0, 0, X_w, Y_w, Z_w, 1, -y_c X_w, -y_c Y_w, -y_c Z_w \end{bmatrix}$$

and $Y_c = \begin{bmatrix} x_c \\ y_c \end{bmatrix}$.

During calibration, $X_c$ and $Y_c$ are known, the unknown $\theta_c$ could be solved by least squares method as follows.

$$\theta_c = (X_c^T X_c)^{-1} X_c^T Y_c \quad (13)$$

The mathematical relationship between the projector coordinate system and the world coordinate system also follows the rule of pinhole camera model that is formulated as follows.

$$Z_p \begin{bmatrix} x_p \\ y_p \\ 1 \end{bmatrix} = \begin{bmatrix} m_{11}^{wp}, m_{12}^{wp}, m_{13}^{wp}, m_{14}^{wp} \\ m_{21}^{wp}, m_{22}^{wp}, m_{23}^{wp}, m_{24}^{wp} \\ m_{31}^{wp}, m_{32}^{wp}, m_{33}^{wp}, m_{34}^{wp} \end{bmatrix} \begin{bmatrix} X_w \\ Y_w \\ Z_w \\ 1 \end{bmatrix} \quad (14)$$

where $(x_p, y_p)$ is the projector coordinate. To decrease the number of unknown parameters, we also let $m_{34}^{wp} = 1$. Eq. (14) is then expanded in the same way as Eq. (4) into three equations.

Since the phase pattern only changes in the $y$ direction, only $y_p$ will be used for 3D measurement. Thus, only the following equation is used.

$$m_{21}^{wp} X_w + m_{22}^{wp} Y_w + m_{23}^{wp} Z_w + m_{24}^{wp} = y_p m_{31}^{wp} X_w + y_p m_{32}^{wp} Y_w + y_p m_{33}^{wp} Z_w + y_p \quad (15)$$

Eq.(15) could be rewritten into the following format:

$$X_p \theta_p = Y_p \quad (16)$$

where $X_p = [X_w, Y_w, Z_w, 1, -y_p X_w, -y_p Y_w, -y_p Z_w]$, $\theta_p = [m_{21}^{wp}, m_{22}^{wp}, m_{23}^{wp}, m_{24}^{wp}, m_{31}^{wp}, m_{32}^{wp}, m_{33}^{wp}]^T$ and $Y_p = [y_p]$.

During calibration, $X_p$ and $Y_p$ are known, the unknown $\theta_p$ could be solved by least squares method as follows.

$$\theta_p = (X_p^T X_p)^{-1} X_p^T Y_p \quad (17)$$

After $\theta_c$ and $\theta_p$ are obtained, they are used for 3D measurement by the following equation.

$$\begin{bmatrix} X_w \\ Y_w \\ Z_w \end{bmatrix} = H^{-1} \begin{bmatrix} x_c - m_{14}^{wc} \\ y_c - m_{24}^{wc} \\ y_p - m_{14}^{wp} \end{bmatrix} \quad (18)$$

where

$$H = \begin{bmatrix} m_{11}^{wc} - x_c m_{31}^{wc}, m_{12}^{wc} - x_c m_{32}^{wc}, m_{13}^{wc} - x_c m_{33}^{wc} \\ m_{21}^{wc} - y_c m_{31}^{wc}, m_{22}^{wc} - y_c m_{32}^{wc}, m_{23}^{wc} - y_c m_{33}^{wc} \\ m_{11}^{wp} - y_p m_{21}^{wp}, m_{12}^{wp} - y_p m_{22}^{wp}, m_{13}^{wp} - y_p m_{23}^{wp} \end{bmatrix}$$

As can be seen, the non-linear mapping function between the 3D world coordinate $(X_w, Y_w, Z_w)$ and $(x_c, y_c, y_p)$ could be determined robustly by the least squares method. For understanding simplicity, we denote $(x_c, y_c, y_p)$ as the phase pattern coordinate, where $(x_c, y_c)$ is obtained based on the index of the acquired image and $y_p$ is obtained by the phase measuring methods. Here, $y_p$ denotes the phase variation with reference to the designed phase pattern. If the coordinates $(x_c, y_c, y_p)$ of all the points in the phase pattern could be determined robustly, the object could be measured robustly with full-resolution. Hence, the major challenge of structured light surface imaging technology lies in how to robustly extract the full-resolution pattern coordinate from the acquired single image or multiple images.

## 3 THE CONCEPT OF PSP

The concept of phase sampling profilometry (PSP) is to measure the sampled phase variation in the spatial domain within a single image by line segmentaion and line clustering algorithms. The full-resolusion phase signal is recovered by Fourier transform and Nyquist-Shannon sampling theory. The designed PSP pattern is formulated as:

$$I_s(x, y) = \left[ I_0 e^{(2\pi x)} \right] \sum_{n=-\infty}^{\infty} \delta(s - nT_S) \quad (19)$$

where $I_0$ is a constant. $T_S$ denotes the Nyquist sampling period. $\delta_T(t) = \sum_{n=-\infty}^{\infty} \delta(t - nT_S)$ is the periodic sampling signal.

The PSP phase pattern was projected onto the measured object, the deformed PSP phase pattern was acquired by a CCD camera and formulated as:

$$\bar{I}_s(x, y) = \gamma(x, y) \left[ I_0 e^{(2\pi x + \varphi(x,y))} \right] \sum_{n=-\infty}^{\infty} \delta(x - nT_S) \quad (20)$$

where $\gamma(x, y)$ is the non-uniform reflectivity distribution and $\varphi(x, y)$ is the phase variation caused by the shape of the object. As can be seen, the phase in the designed PSP pattern only changes in the vertical direction and only the dimension in the vertical direction is useful for phase variation calculation. The CCD captured one-dimension phase signal in the vertical direction is formulated as:

$$\bar{f}_s(t) = \gamma(t) \left[ I_0 e^{(2\pi t + \varphi(t))} \right] \sum_{n=-\infty}^{\infty} \delta(t - nT_S) \quad (21)$$

The reflectivity modified 1D phase signal $f(t)$ is denoted as:

$$f(t) = \gamma(t) \left[ I_0 e^{(2\pi t + \varphi(t))} \right] \quad (22)$$

The sampling impulse signal $\delta_T(t)$ is formulated as:

$$\delta_T(t) = \sum_{n=-\infty}^{\infty} \delta(t - nT_S) \quad (23)$$

The Fourier transform of the sampling impulse signal $\delta_T(t)$ is formulated as:

$$\Delta_T(w) = w_s \sum_{n=-\infty}^{\infty} \delta(w - nw_s) \quad (24)$$

where $w_s$ is the sampling angle frequency. The Fourier transform of the CCD captured one-dimension PSP signal is formulated as:

$$F_s(w) = \frac{1}{2\pi} F(w) * \Delta_T(w) \quad (25)$$

where $F(w)$ is the Fourier transform of the 1D reflectivity function $f(t)$.

A rectangle window function is used to obtain the spectrum of the reflectivity modified 1D phase signal $\bar{f}(t)$ from $F_s(w)$ as follows:

$$\bar{F}(w) = F_s(w) H(w) \quad (26)$$

where $H(w)$ is the rectangle window function and it is formulated as:

$$H(w) = \begin{cases} T_s & |w| \leq w_s/2 \\ 0 & |w| > w_s/2 \end{cases} \quad (27)$$

In the time domain, the obtained reflectivity modified one dimension phase pattern $\bar{f}(t)$ is formulated as:

$$\bar{f}(t) = \sum_{n=-\infty}^{\infty} f(nT_s) \frac{\sin w_m(t-nT_s)}{w_m(t-nT_s)} \quad (28)$$

where $w_m$ is the maximum angle frequency of $F(w)$. According to Shannon sampling theory, if $w_s \geq 2w_m$, then $\bar{f}(t) = f(t)$, i.e. the reflectivity modified 1D phase signal could be recovered perfectly.

Since $f_s(t)$ has been determined by image processing algorithms, $f(t)$ could also be determined based on Eq. (28). From Eq. (4), it is seen that the mapping relationship between $f(t)$ and $\varphi(t)$ is nonlinear and it could be determined by the least squares method. Hence, $f(t)$ could be used directly as the phase variation $y_p$ for 3D measurement.

## 4 EXPERIMENTAL RESULTS

We use a simulated phase pattern to demonstrate how PSP works to recover the sampled phase signal. Fig. 2 (a) shows the PSP phase pattern generated by Eq. (19) with $\varphi(x,y)=0$ and $\gamma(x,y)=1$. Fig. 2 (b) shows the 1D PSP phase signal $f_s(t)$ generated by Eq. (21) with $\varphi(t)=0$ and $\gamma(t)=1$, where the continuous phase signal $f(t)$ is denoted by the green line and the sampling impulse signal $\delta_T(t)$ is denoted by the red lines. Fig. 2 (c) shows the Fourier spectrum of the 1D PSP phase signal $f_s(t)$. Fig. 2 (d) shows the recovered continuous phase signal $\bar{f}(t)$ by the proposed method and the time domain spline interpolation method with reference to the ground truth phase signal, where the green line denotes the ground truth phase signal, the blue line denotes the phase signal recovered by the spline interpolation method[19-21] and the red line denotes the phase signal recovered by the proposed method. For the PSP pattern demonstrated in Fig. 3, the sampling period $T_s$ is 17 pixels. The average phase recovering error of the spline interpolation method is 1.472 while the average phase recovering error of the proposed method is 2.4216. In Fig. 3, we demonstrate another PSP pattern with the sampling rate $T_s = 27$. In Fig. 3 (d), the recovered phase signal by the proposed method matches the ground truth phase signal completely and the average phase recovering error of the proposed method decreases to 0. On the contrary, the average phase recovering error of the spline interpolation method increases to 4.8146. From the two demonstrations shown in Fig. 2 and Fig. 3, it is seen that the sampling rate is very important for PSP.

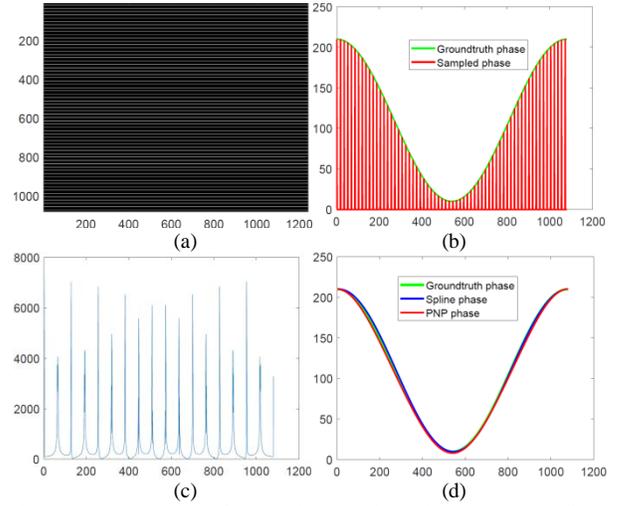

Fig. 2. Demonstration of recovering the phase signal with sampling rate $T_s = 17$ (a) The generated PNP phase pattern sampled with $T_s = 17$; (b) The 1D PNP phase signal sampled with $T_s = 17$; (c) Fourier spectrum of the sampled phase signal; (d) Recovered phase signal.

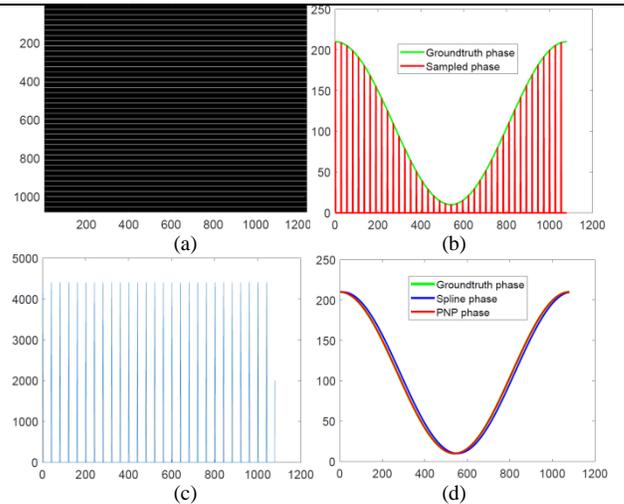

Fig. 3. Demonstration of recovering the phase signal with sampling rate $T_s = 27$ (a) The generated PNP phase pattern sampled with $T_s = 27$; (b) The 1D PNP phase signal sampled with $T_s = 27$; (c) Fourier spectrum of the sampled phase signal; (d) Recovered phase signal.

## 5 CONCLUSION

In this letter, the concept of PSP is introduced and its underlying theory is explained. With the Nyquist sampling rate, PSP could recover the full-resolution phase accurately with only one single image. Thus, PSP is more suitable for measuring the moving objects than many other techniques. Experimental results showed that the proposed frequency domain phase recovering method is more accurate than the time domain phase recovering method. In addition, the loose requirement of the sampling rate by the frequency domain phase recovering method makes the unsupervised and robust line segmentation and clustering possible for many computer vision applications.